\documentclass[prl,reprint,amssymb,showpacs,aps]{revtex4-1}
\usepackage{graphicx}
\usepackage{amsmath}
\usepackage{braket}
\usepackage{bm}

\begin{document}

\title{Magnetism in Rare Earth Quasicrystals: RKKY Interactions and Ordering}

\author{Stefanie Thiem}
\author{J. T. Chalker}
\affiliation{Theoretical Physics, University of Oxford, 1 Keble Road, Oxford OX1 3NP, United Kingdom}
\date{\today}

\begin{abstract}
We study magnetism in simple models for rare earth quasicrystals, by considering Ising spins on a quasiperiodic tiling, coupled via RKKY interactions. Computing these interactions from a tight-binding model on the tiling, we find that they are frustrated and strongly dependent on the local environment. Although such features are often associated with spin glass behaviour, we show using Monte Carlo simulations that the spin system has a phase transition to a low-temperature state with long-range quasiperiodic magnetic order.
\end{abstract}


\maketitle


Understanding how the magnetic properties of complex materials arise from their atomic structure represents one of the fundamental challenges of condensed matter physics. In the context of quasicrystals \cite{PhysRevLett.1984.Shechtman} this leads to the question of how their magnetism \cite{PhilMag.2008.Hippert, MagProp.2013.Stadnik} is influenced by their unusual electronic properties, which include a pseudogap at the Fermi energy \cite{PhysRevLett.1991.Fujiwara, PhysRevLett.1992.Hafner}, multifractal wave functions \cite{PhysRevB.1987.Kohmoto} and anomalous electronic transport \cite{PhysicalProperties.1999.Stadnik, JPhysJap.1987.Hiramoto}. Magnetic rare earth (R) quasicrystals in particular serve to motivate simple models, as they
contain well-defined local moments at concentrations of 5-10\% interacting via long-range RKKY interactions \cite{PhysRevB.1999.Fisher,NatureMat.2013.Goldman}. Examples of these quasicrystals \cite{MagProp.2013.Stadnik} include the icosahedral i-ZnMgR \cite{PhysRevB.1999.Fisher} and i-AgInR  \cite{PhilMagLett.2002.Guo} materials, as well as some decagonal d-ZnMgR materials \cite{PhilMagLett.1998.Sato} and the recently discovered binary phases i-RCd \cite{NatureMat.2013.Goldman}.

In this paper we take a two-step theoretical approach to studying such systems. First, we compute RKKY interactions from a tight-binding Hamiltonian defined on a two-dimensional quasiperiodic tiling. Second, we examine the statistical mechanics of Ising spins with these interactions, using extensive Monte Carlo simulations. While there has been much previous work studying spin models for magnetic quasicrystals \cite{JStatPhys.1986.Godreche,PhysRevLett.2003.Wessel,PhysRevLett.2003.Vedmedenko,PhysRevLett.2004.Vedmedenko,PhilMag.2006.Vedmedenko,JNonCrys.2004.Matsuo,JMagMagMat.2002.Matsuo, PhilMag.2006.Matsuo}, it has been based on choices of interaction (nearest neighbour exchange \cite{JStatPhys.1986.Godreche,PhysRevLett.2003.Wessel}, dipolar and further neighbour exchange \cite{PhysRevLett.2003.Vedmedenko,PhysRevLett.2004.Vedmedenko,PhilMag.2006.Vedmedenko}, or RKKY interactions with a form taken from periodic systems \cite{JNonCrys.2004.Matsuo,JMagMagMat.2002.Matsuo, PhilMag.2006.Matsuo}) that do not incorporate the unique coupling of electronic to magnetic properties that is to be expected in quasicrystals. Our model is designed to address this aspect of the physics in a simple way.

We find substantial differences between RKKY interactions on a quasiperiodic tiling and those in simple metals  \cite{PhysRev.1954.Ruderman}. While both ferromagnetic and antiferromagnetic interactions arise in the tiling, they have no well-defined spatial period since there is no Fermi wavevector. And as there is no translational invariance, the RKKY coupling between pairs of sites is a function not only of their separation, but is also strongly dependent on the local environment. One might expect that the combination of frustration and aperiodicity would lead to spin freezing \cite{RevModPhys.1986.Binder}, and in other settings quasiperiodic systems are known to behave like random ones \cite{PhysRevLett.1982.Fishman}. Our results, however, exclude canonical spin glass behaviour, via the scaling of domain wall energy with length. We demonstrate that instead there is quasiperiodic spin order at low temperature, by examining the Fourier transform of spin configurations. 

\begin{figure}[b!]
    \centering
    \includegraphics[width=0.95\columnwidth]{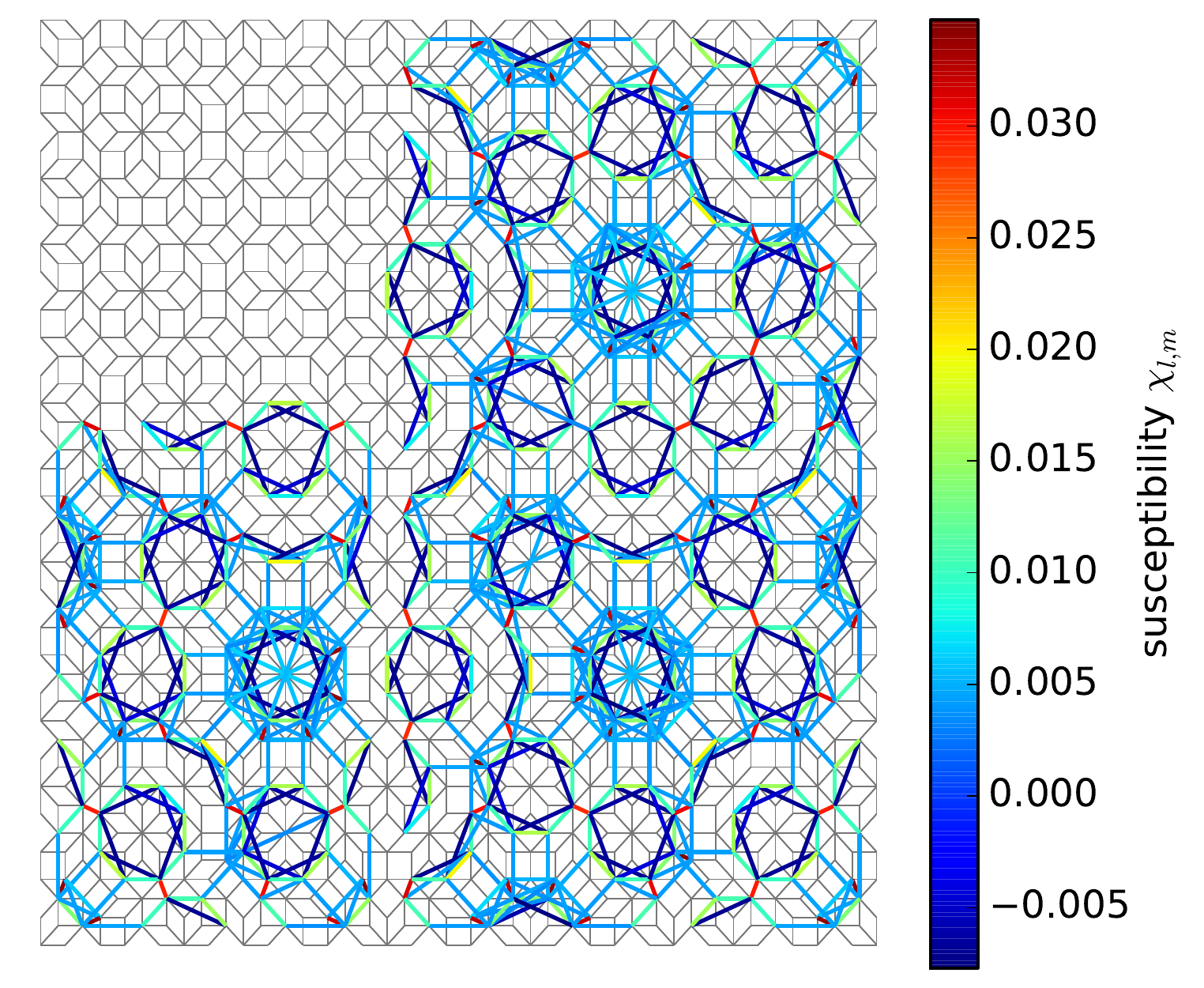}
  \caption{Ammann Beenker tiling (top left corner) and RKKY interactions $\chi_{l,m}$ (threshold $|\chi_{l,m}| > 0.004$) for magnetic moments at sites with coordination number $z=4$ and $E_\mathrm{F} = 1.95$.}
\label{fig:tilings}
\end{figure}


We model conduction electrons by a pure hopping Hamiltonian $H_\textrm{el} = \sum_{\langle l,m \rangle} \ket{l}\bra{m}$ with one orbital per site and equal hopping amplitudes between all nearest neighbours of a quasiperiodic tiling (see fig.~\ref{fig:tilings}). This approach takes the multifractallity of the electronic states into account \cite{JPhys.1998.Rieth,PhilosMag.2011.Trambly}.  Introducing Ising spins $\sigma_l$ at some of the sites $l$ of the tiling, the spin Hamiltonian has the form  $H_\mathrm{RKKY} = \lambda^2 \chi_{l,m} \sigma_l \sigma_m$. Here, $\lambda$ represents the coupling of the local moment to conduction electrons, and the local susceptibility  for zero temperature and Fermi energy $E_\mathrm{F}$ is (\cite{JPhysFrance.1994.Jagannathan}; for a review see \cite{Crystals.2013.Power})
\begin{equation}
    \label{equ:susceptibility}
    \chi_{l,m} = \frac{1}{\pi} \int_{-\infty}^{\infty} \Im\left[ G_{l,m} G_{m,l} \right] \mathrm{sign} (E - E_\mathrm{F}) \mathrm{d}E ,
\end{equation}
where $G_{l,m} \equiv \bra{l} G \ket{m}$ is a matrix element of the retarded Green function $G\equiv [E+i0-H_\textrm{el}]^{-1}$ for the conduction electrons.

\begin{figure}[t!]
    \centering
    \includegraphics[height=4.05cm]{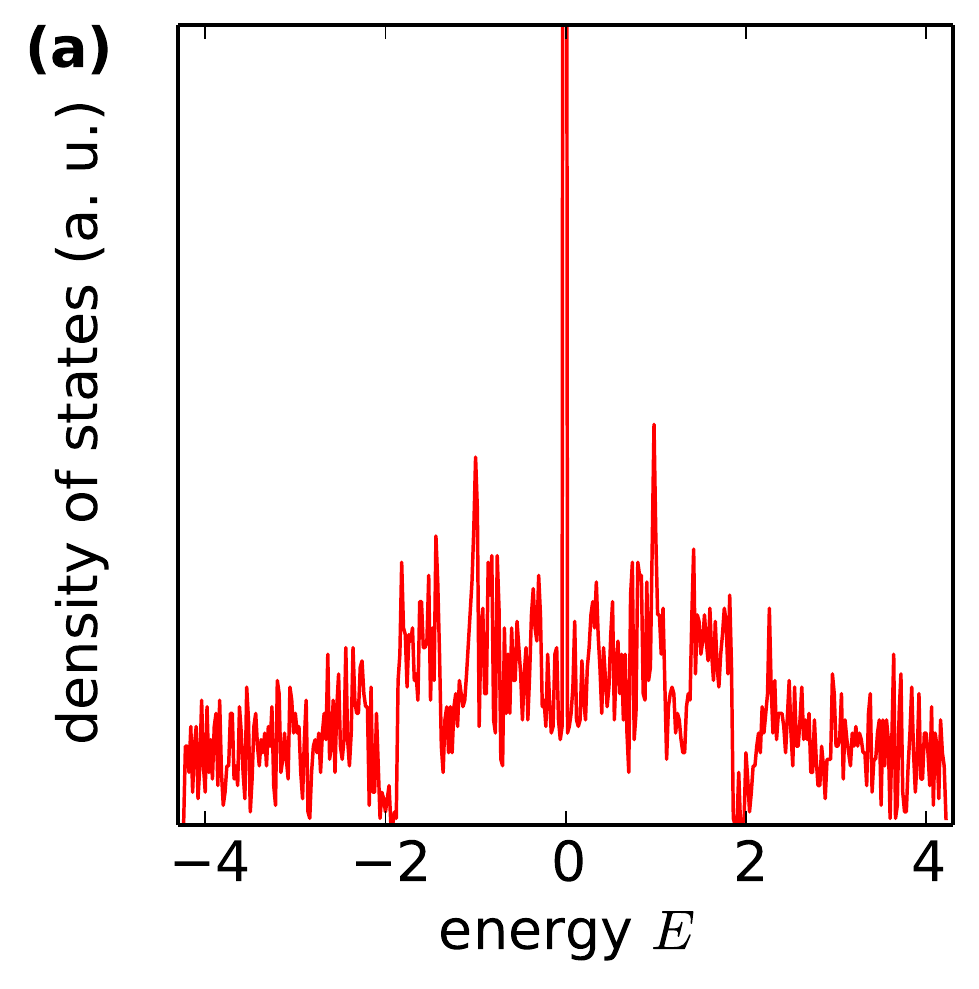}\hfil
    \includegraphics[height=4.1cm]{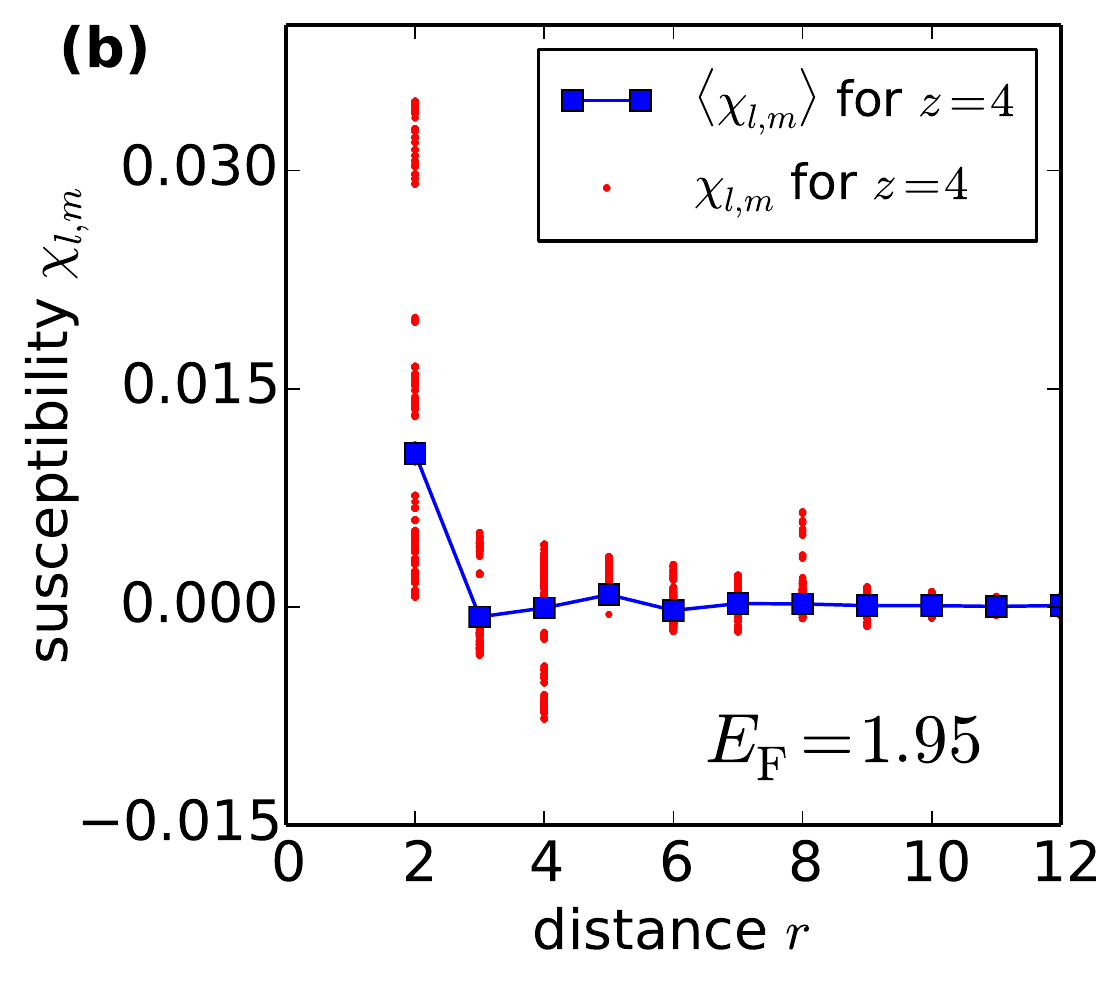}\\
    \includegraphics[height=4.1cm]{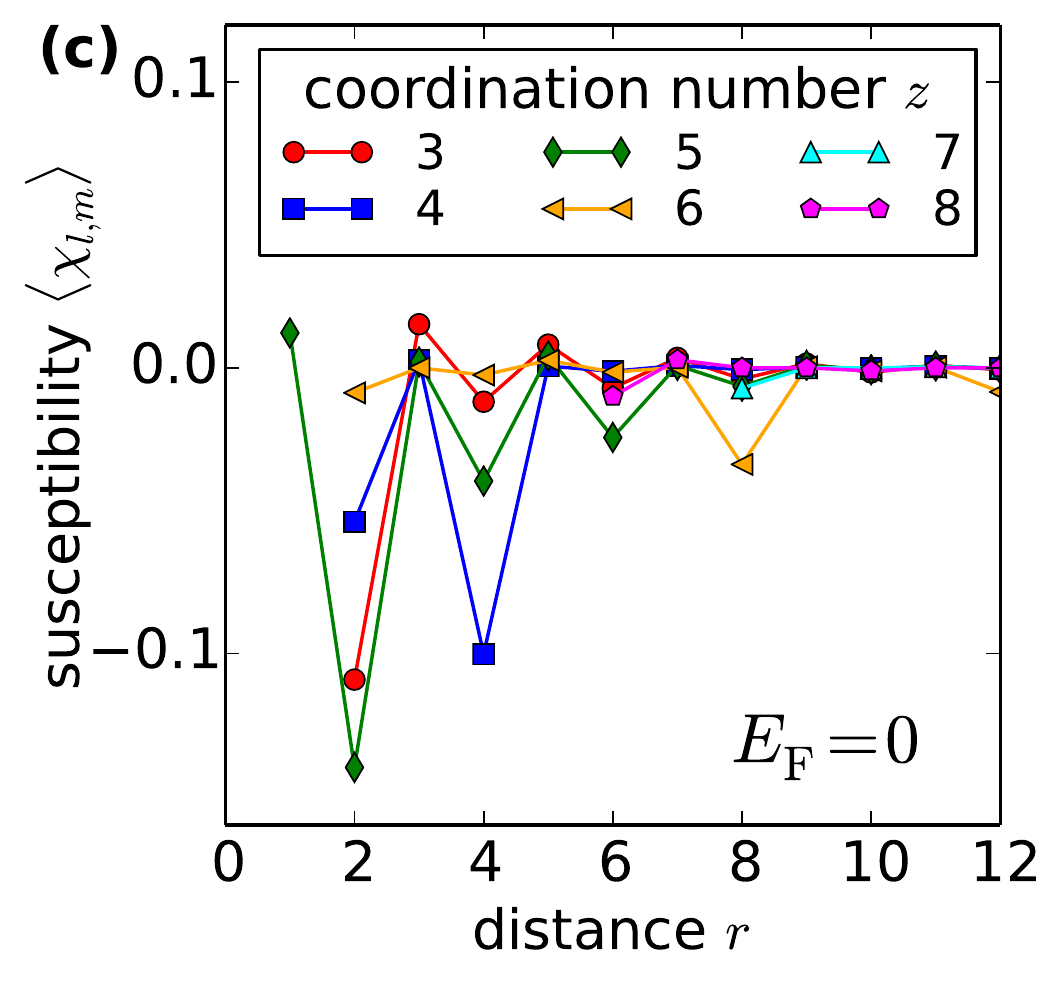}\hspace{-0.2cm}
    \includegraphics[height=4.1cm]{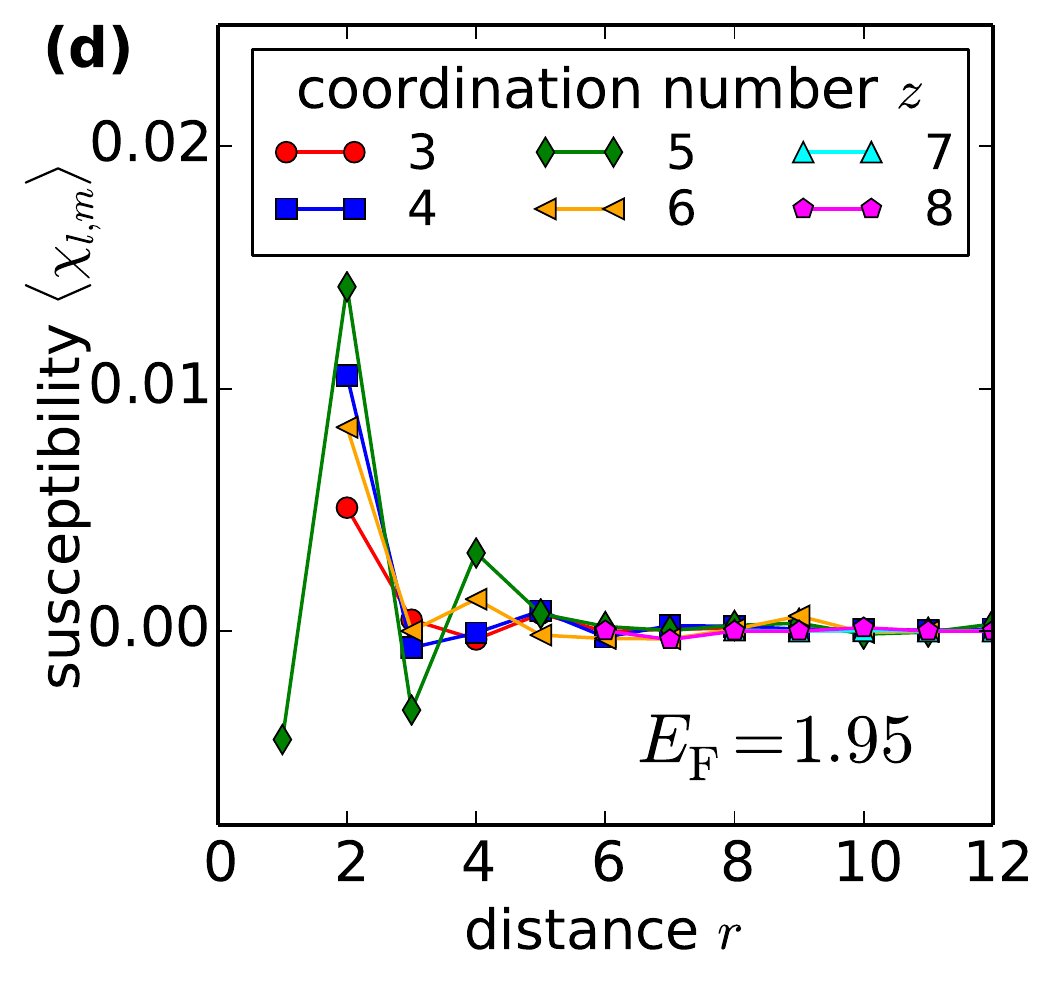}
\caption{Results for Ammann Beenker tiling with 1393 sites: (a) DOS of the conduction electrons, (b) distribution of local susceptibility $\chi_{l,m}$ for $z=4$ and $E_\mathrm{F} = 1.95$, and average local susceptibility for different $z$ for (c) $E_\mathrm{F} = 0$ and (d) $E_\mathrm{F} = 1.95$.}
\label{fig:chi-octagonal}
\end{figure}

To compute the Green function we use a continued fraction expansion, which has been employed previously for quasiperiodic tilings \cite{JPhysFrance.1994.Jagannathan, PhysRevB.1987.Kumar}. It has advantages over the alternatives: direct diagonalization is limited to relatively small sample sizes, while an expansion of the Green function in Chebyshev polynomials, employed in Ref.~\cite{PhysRevB.1999.Roche}, has the drawback of producing artefacts for the spiky density of states (DOS) that is a typical feature of quasicrystals (see fig.~\ref{fig:chi-octagonal}a).
 
The computation of the off-diagonal Green function elements $G_{l,m}$ in eq.~\eqref{equ:susceptibility} can be reduced to the evaluation of three diagonal elements using the identity
$
2\,G_{l,m} =  (1+\mathrm{i})  \bra{\Psi^+} G \ket{\Psi^+} + (-1+\mathrm{i}) \bra{\Psi^-} G \ket{\Psi^-} - 2 \mathrm{i} \bra{\Psi^\mathrm{im}} G \ket{\Psi^\mathrm{im}} ,
$
where $\ket{\Psi^\pm} =\frac{1}{\sqrt{2}}  (\ket{l}  \pm \ket{m})$ and $\ket{\Psi^\mathrm{im}} = \frac{1}{\sqrt{2}}(\ket{l} + \mathrm{i} \ket{m})$. Each of the diagonal elements $G_{XX}\equiv \bra{\Psi^X} G \ket{\Psi^X} $ can be expanded as a continued fraction
\begin{equation}
G_{XX} = \frac{1}{E - a_1 - \frac{b_1}{E - a_2 - \frac{b_2}{E - a_3 - \ldots }}} .
\end{equation}
The terms $a_n$ and $b_n$ are obtained by the tridiagonalization of the Hamiltonian $H_\textrm{el}$. Finite system size limits the number of possible tridiagonalization steps, and to take the infinite environment into account, we approximate the tiling beyond a certain distance (roughly the size of the periodic approximant) by an average structure \cite{JPhysC.1975.Haydock}. For an infinite system with a gapless and symmetric DOS we can use $a_\infty = 0$ and $b_\infty = E_\mathrm{max}^2 / 4$. The continued fraction expansion yields a smoothed version of the Green functions and the susceptibility \cite{JPhysFrance.1994.Jagannathan}, which is reasonable because disorder and finite temperatures also lead to a smoothing of the sharp peaks of the Green functions in quasiperiodic tilings.

The approach can be applied to general tight-binding models. For a particular model we have to specify  three features: (i) the tiling; (ii) the positions of the magnetic sites; and (iii) the value of $E_\mathrm{F}$. Here we study the Ammann Beenker tiling (see fig.~\ref{fig:tilings}), which models the structure of octagonal quasicrystals \cite{DCGeom.1992.Ammann}. It possesses 6 different local environments with 3 to 8 nearest neighbours. We consider the 4th and 5th approximant with 1393 and 8119 sites respectively. To take into account the high degree of order in stable quasicrystals \cite{PhysicalProperties.1999.Stadnik}, we place magnetic moments at the subset of sites with a specified coordination number $z$. 

Our results for the local susceptibility $\chi_{l,m}$ are shown in Figs.~\ref{fig:tilings} and \ref{fig:chi-octagonal} \footnote{Results given in a previous version of this paper (arXiv:1407.5868v1) were affected by a coding error in the calculation using the continued fraction expansion.
The corrected results presented in this revised version are different in detail but our physical conclusions are unchanged.}. There is a clear dependence of the susceptibility on $E_\mathrm{F}$ and on the local environment, where bonds of similar strength occur at the same local pattern of the tiling as shown in fig.~\ref{fig:tilings}. This is expected since $\chi_{l,m}$ is generally large if the local DOS at $E_\mathrm{F}$ is large, and the local DOS is known to vary significantly with local environment \cite{JPhysFrance.1994.Jagannathan}. Correspondingly, interactions are also smaller when $E_\mathrm{F}$ lies in a pseudogap. In general, we find small interaction strengths for local environments with high coordination number, and oscillations between ferromagnetic and antiferromagnetic interactions as a function of site separation $r$, measured along the shortest bond path (see fig.~\ref{fig:chi-octagonal}). This behaviour differs significantly from that in periodic systems with a spherical Fermi surface, where interactions oscillate within a power-law envelope that varies as $r^{-d}$. Even with parameters for which the average $\langle \chi_{l,m} \rangle$ is reasonably well described by this power law, individual interactions in quasiperiodic tilings can be considerably larger, and we usually observe a hierarchy of interaction strengths at each $r$ (fig.~\ref{fig:chi-octagonal}b), which results in a quasi-random contribution to the magnetic interactions due to the many local environments in a quasicrystal.

We focus on two specific values: $E_\mathrm{F} = 0$ and $E_\mathrm{F} = 1.95$.  The first is an example of a bipartite lattice at half-filling, and RKKY interactions are therefore antiferromagnetic ($\chi_{l,m}>0$) between sites on opposite sublattices and ferromagnetic ($\chi_{l,m}<0$) on the same sublattice \cite{PhysRevB.2007.Saremi}, as indeed found in fig.~\ref{fig:chi-octagonal}c. The second choice locates the Fermi energy at the minimum of the most prominent pseudogap in the DOS, a typical feature of quasicrystals. In this case $\chi_{l,m}$ oscillates much less with $r$.


In the second part of this paper we study the statistical mechanics of the spin Hamiltonian $H_\mathrm{RKKY}$. We use Monte Carlo simulations with parallel tempering, employing multiple copies of the system at different temperatures $T_i$ to reduce correlation times \cite{PhysRevLett.1986.Swendsen, EurPhysLett.1992.Marinari}; simulation parameters will be reported in detail elsewhere \cite{ToAppear.2014.Thiem}. Important observables are the energy $E =  \sum_{l,m} \chi_{l,m} \sigma_l \sigma_m $ and the magnetization  $M  =  \sum_{l=1}^{N} \sigma_l $. Indications of a phase transition may come from the heat capacity per spin $C = \frac{1}{NT^2} \left[ \langle E^2\rangle - \langle E\rangle^2 \right]$ and the susceptibility $\chi = \frac{1}{NT}\left[ \langle M^2\rangle - \langle M\rangle^2 \right]$. To probe a hidden antiferromagnetic order we also compute the magnetization $M_\mathrm{gs} = \sum_{l=1}^N \xi_l \sigma_l$ compared to the ground state configuration $\{\xi_l\}$ and the associated order parameter susceptibility $\chi_\mathrm{op} = \frac{1}{NT}\left[ \langle M_\mathrm{gs}^2\rangle - \langle |M_\mathrm{gs}|\rangle^2 \right]$.
In addition, some quantities originally introduced for the study of spin glasses are useful as probes that are sensitive to a variety of ordering patterns. Specifically, we consider the overlap $q = \frac{1}{N} \sum_{l=1}^{N}  \sigma_l^1 \sigma_l^2 $, obtained from the simulations of two independent replicas with the same interactions and temperature. We compute the Binder cumulant $B_\mathrm{SG} = \frac{1}{2} (3 - \langle q^4\rangle / \langle q^2 \rangle^2 )$, which is a good tool to identify order, whether it is (anti)ferromagnetic or spin-glass-like, as is clear from a consideration of limiting cases. At high temperature, when spins are uncorrelated, we expect $q\sim N^{-1/2}$ and $B_\mathrm{SG}=0$. Conversely, at low temperature, if only a pair of ground states related by a global spin inversion are accessible, one has $q=\pm 1$ and $B_\mathrm{SG}=1$. 


We have carried out simulations for 18 different parameter sets on the Ammann Beenker tiling, varying $E_\mathrm{F}$ and the local environment  of magnetic moments. In all cases we find the same qualitative picture of a phase transition to a state with local antiferromagnetic correlations and broken Ising symmetry. We show below that this low-temperature state has quasiperiodic N\'eel order. 

\begin{figure}
    \includegraphics[width=\columnwidth]{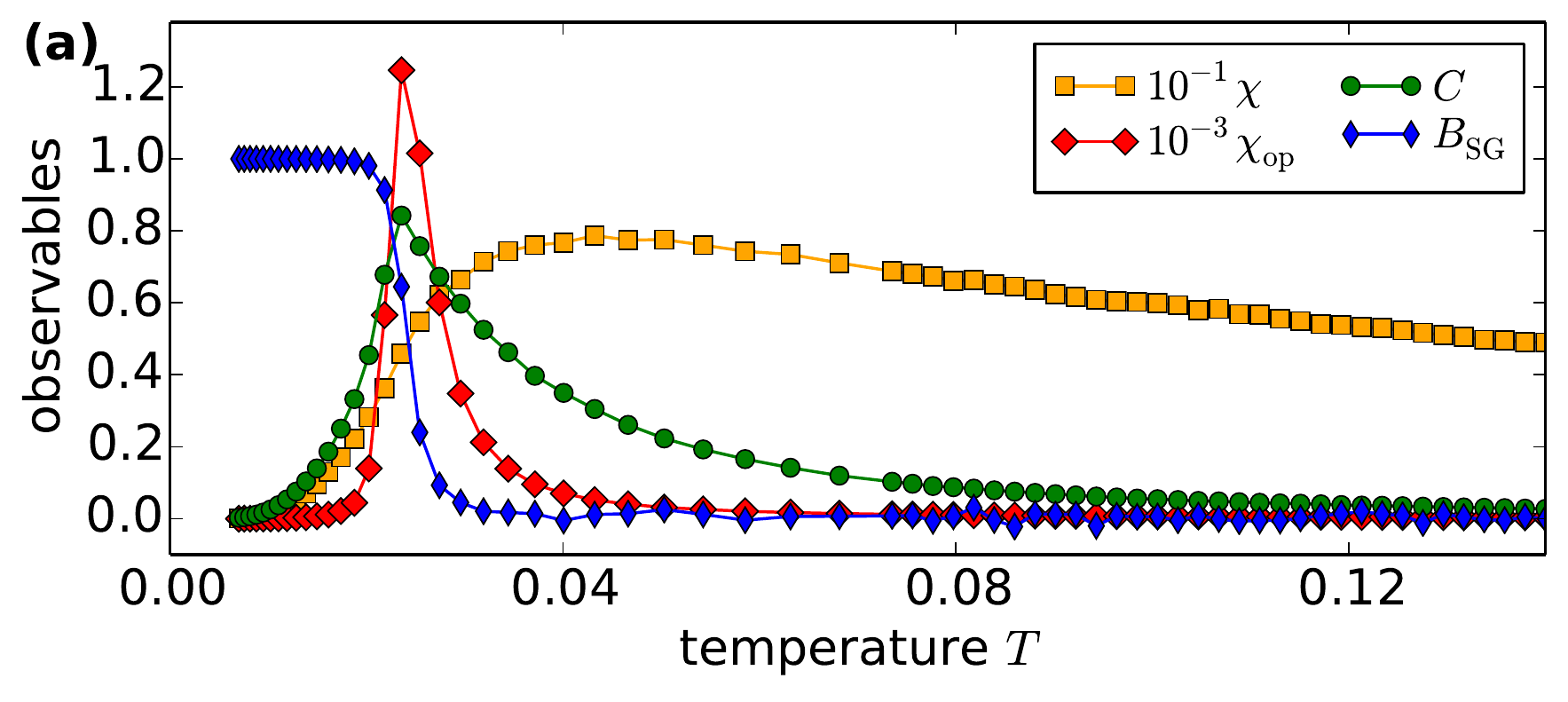}
    \includegraphics[width=\columnwidth]{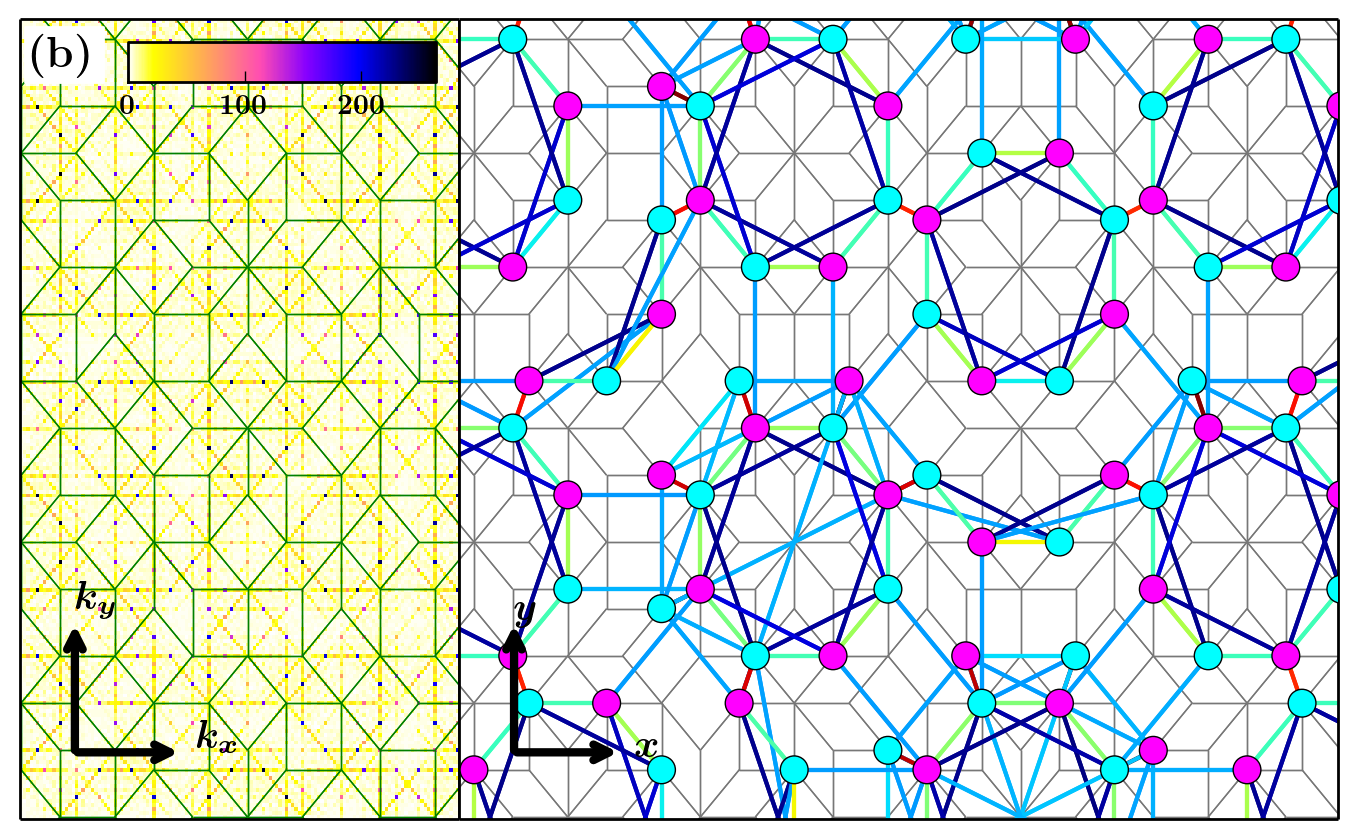}
    \includegraphics[width=\columnwidth]{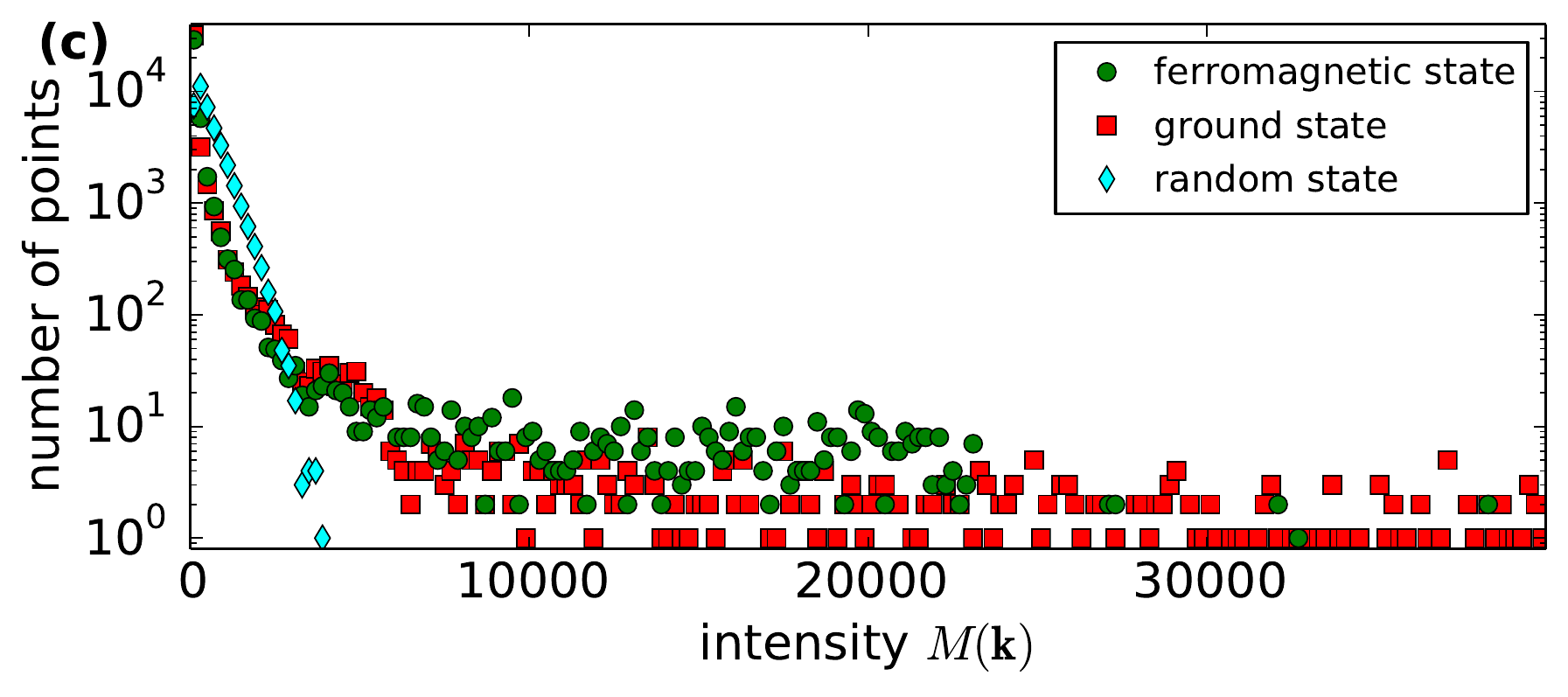}
\caption{Simulation results for the Ammann Beenker tiling with $z=4$ and $E_\mathrm{F} = 1.95$: (a) different observables; (b) lowest energy spin configuration (red and blue circles for up and down spins) with strongest bonds according to fig.~\ref{fig:tilings} (right) and corresponding diffraction pattern amplitude $\sqrt{M(\mathbf{k})}$ with overlaid Ammann Beenker tiling (left);  (d) intensity distribution in diffraction patterns of the ground state, the ferromagnetic state, and a random state.} 
\label{fig:pt-octagonal}
\end{figure}

To illustrate the transition to a quasiperiodic N\'{e}el state, we present in fig.~\ref{fig:pt-octagonal} the simulation results for $N = 478$ Ising spins at sites with coordination number $z=4$ on the approximant of fig.~\ref{fig:tilings}, taking $E_\mathrm{F} = 1.95$ and $\lambda=1$. The heat capacity $C$ and susceptibility $\chi$ have peaks near $T_\mathrm{f} \approx 0.03$, and the Binder cumulant $B_\mathrm{SG}$ approaches 1 below this temperature. This suggests that the system is paramagnetic for $T > T_\mathrm{f} $, and that at $T_\mathrm{f} $ a macroscopic fraction of  spins lock together into a state with long-range rigidity. The peak in the order parameter susceptibility  $\chi_\mathrm{op}$ at $T_\mathrm{f}$ is associated with the spontaneous breaking of the global Ising symmetry of the spin model when spins order into a low-temperature phase with zero magnetization $M$ but non-zero $M_{\rm gs}$.

A key question is what type of order characterises the low-temperature phase. Short-distance correlations are readily apparent in the ground state spin configuration (fig.~\ref{fig:pt-octagonal}b right), which contains many octagons with an antiferromagnetic spin configuration forming on the 8-fold stars of the Ammann Beenker tiling. Long range order results from a tendency of neighbouring clusters to anti-align due to strong antiferromagnetic bonds between the clusters. The nature of this order is revealed to be quasiperiodic by the diffraction intensity $M(\mathbf{k}) = \left|\sum_l e^{2\pi \mathrm{i}\cdot \mathbf{k} \mathbf{r}_l} \sigma_l \right|^2$ of the low-temperature spin configurations. In contrast to periodic systems, quasicrystals do not posses a Brillouin zone. Instead, long-range quasiperiodic order leads to Bragg peaks in a pattern that forms a reciprocal-space Ammann Beenker tiling with a length scale of $1/(2b)$ in units of the real-space bond length $b$ \cite{ActaCrysA.1988.Wang, CrysQuasicrystals.Steurer}. Numerical results for $M(\mathbf{k})$ in fig.~\ref{fig:pt-octagonal}b (left) indeed show a pattern of high-intensity peaks which can be overlaid by an Ammann Beenker tiling with the expected length scale. A possible concern is whether the quasiperiodic site locations by themselves might be sufficient to generate these features in the diffraction pattern, without magnetic order. To test this we have computed the diffraction pattern for a random spin configuration on the tiling: its amplitude fluctuations are Gaussian and its intensity distribution (fig.~\ref{fig:pt-octagonal}c) does not show the high intensity peaks that are present for the ground state and for a ferromagnetic one. This clearly indicates that the observed pattern is due to the long-range magnetic order and not due to the atomic order.

\begin{figure}
    \includegraphics[width=\columnwidth]{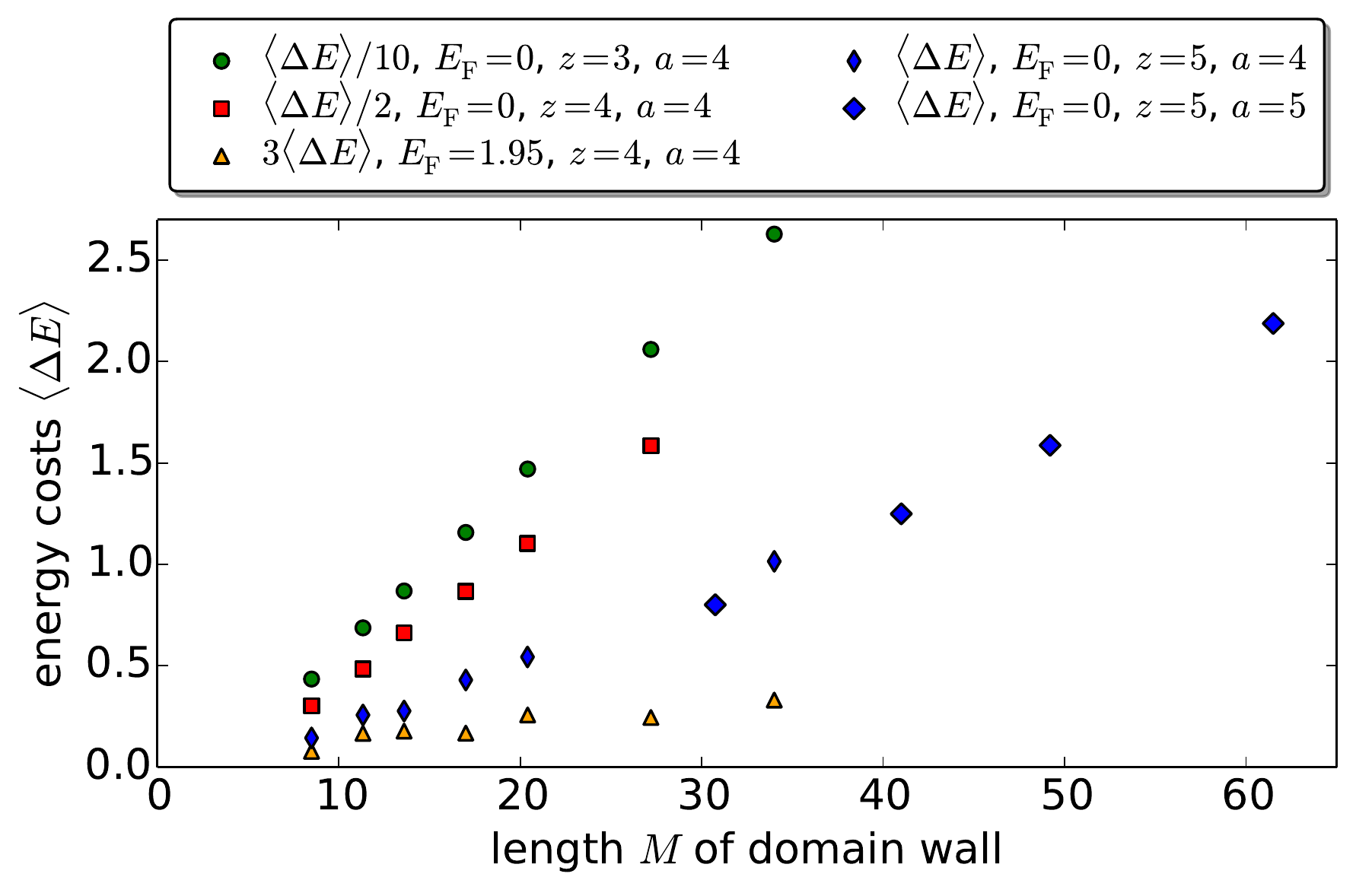}
\caption{Scaling of the energy costs of a domain wall with its length for different parameter sets on the Ammann Beenker tiling.} 
\label{fig:pt-ars}
\end{figure}

As another probe for the long-range order of the low temperature state, we also investigate the ground state energy cost of introducing a domain wall, and its scaling with system size, an approach that has a long history in the study of spin glasses \cite{JPhysC.1984.Bray, PhysRevB.1998.Matsubara, PhysRevLett.2002.Carter}. The cost should diverge with system size in an ordered state that is stable at finite temperature. The fact that it instead falls towards zero in two-dimensional models of short-range spin-glasses with a continuous distribution of interaction strengths is evidence against a finite freezing temperature in those systems. To examine this behaviour in our models we cut a strip of width $M$ (in units of the bond length $b$) from an approximant of length $L$. To introduce a domain wall across the strip, we compare the effects of periodic and antiperiodic boundary conditions along the length, with free boundary conditions at the sides, by determining the lowest energies $E_\mathrm{p}$ and $E_\mathrm{ap}$ in each case using parallel tempering. The energy cost of the domain wall is $\Delta E = \left| E_\mathrm{ap} - E_\mathrm{p}\right|$. We average over different strips for each ratio $R = L/M$. The results for the 4th and 5th approximant are shown in fig.~\ref{fig:pt-ars}. We obtain to a good approximation $\Delta E \propto M$. Inspecting individual spin configurations, we find that this reflects the formation of fairly straight domain walls that avoid the strongly-coupled clusters in the system. Our results correspond to a stiffness exponent of $\theta \approx 1$, in stark contrast to the value $\theta<0$ for conventional spin glasses in two dimensions \cite{PhysRevLett.2002.Carter}.

There are two main limitations to our simulations. First, while finite size scaling is a valuable tool for studying phase transitions, it is difficult to employ it using the Ammann Beenker tiling with periodic boundary conditions, as the system size of the successive periodic approximants are in the ratio $\tau = 3+2\sqrt{2}$. For this reason the results described above are restricted to only two system sizes. Second, although dynamic behaviour (specifically the difference between field-cooled and zero field-cooled magnetisation) is an important experimental probe, we cannot access it in our parallel tempering simulations because this method is specifically designed to reduce correlation times.


The model we have introduced has some distinctive features in addition to those arising from ordering. It is notable that RKKY interactions generate strongly coupled spin clusters in all of the considered systems, as illustrated in fig.~\ref{fig:pt-octagonal}b. For a given choice of magnetic site and $E_{\mathrm{F}}$ the most prominent clusters have a fixed form and occur at the same local pattern of the tiling. As anticipated from Conway's theorem that any local pattern of linear dimension $L$ is repeated in a distance $\mathcal{O}(L)$ for each quasiperiodic tiling \cite{SciAm.1977.Gardner}, the strongly coupled spin clusters  repeat frequently. This feature appears consistent with experimental observations of the formation of ordered spin clusters at low temperatures \cite{PhysRevB.2000.Sato, PhysRevB.2006.Sato}.

So far quasiperiodic magnetic order (rather than spin freezing at low temperature) has not been reported for rare earth quasicrystals \cite{MagProp.2013.Stadnik}. Experimental data show a spin-glass-like separation between field-cooled and zero-field-cooled magnetic susceptibility below a freezing temperature $T_{\rm f}$ \cite{PhilMag.2008.Hippert, MagProp.2013.Stadnik, NatureMat.2013.Goldman}. The Weiss temperature $\Theta$ is negative in most cases, indicating predominantly antiferromagnetic interactions, and large values $(5-10)$ of $-\Theta/T_{\rm f}$ imply strong frustration \cite{PhysRevB.1999.Fisher,NatureMat.2013.Goldman}. However, some 1/1 cubic approximants (e.g.\ Cd$_6$R \cite{PhysRevB.1998.Islam}, AuSiGd \cite{JPhysCondMat.2013.Hiroto}) show antiferromagnetic or ferromagnetic order for samples with sufficiently low site disorder of the rare earth atoms. We note that site disorder is very common in quasicrystals, and may be the reason for spin-glass behaviour rather than long-range magnetic order. Moreover, even disorder restricted to non-magnetic sites will influence the couplings between spins. We find in preliminary work that RKKY interactions computed in our model show a high sensitivity to a weak random on-site potential.

 
In summary, we have used tight binding models and spin models defined on quasiperiodic tilings to examine how the unusual electronic properties of quasicrystals influence their magnetism. Despite the occurrence of strong frustration and quasi-random noise for the magnetic interactions, the spin models show a phase transition towards a quasiperiodic N\'eel state with a hidden long-range antiferromagnetic order instead of a canonical spin-glass transition. In addition, we find that RKKY interactions computed from the multifractal electronic states of a quasiperiodic tiling lead to the emergence of small, strongly coupled clusters of spins. 

This research was supported by a Marie Curie Intra European Fellowship within the 7th European Community Framework Programme and by EPSRC Grant No. EP/I032487/1. 

\bibliographystyle{apsrev4-1}
\bibliography{paper}

\end{document}